\documentclass[twocolumn,showpacs,amsmath,amssymb,showkeys]{revtex4}

\usepackage{amssymb}
\usepackage{amsmath}
\usepackage{exscale}

\usepackage{graphics}

\begin{document}

\newcommand{\be}{\begin{equation}}
\newcommand{\ee}{\end{equation}}
\newcommand{\ben}{\begin{eqnarray}}
\newcommand{\een}{\end{eqnarray}}
\newcommand{\nn}{\nonumber \\}
\newcommand{\ii}{\'{\i}}
\newcommand{\pp}{\prime}
\newcommand{\expq}{e_q}
\newcommand{\lnq}{\ln_q}
\newcommand{\quno}{q-1}
\newcommand{\qunoinv}{\frac{1}{q-1}}
\newcommand{\tr}{{\mathrm{Tr}}}
\newcommand{\nd}{{\noindent}}

\title{Delocalization and the semiclassical description of molecular rotation}

\author{S.~Curilef$^1$,  F.~Pennini$^{1,2}$,  A.~Plastino$^2$, and G.L.~Ferri$^3$}



 \affiliation{$^1$Departamento de F\'{\i}sica, Universidad Cat\'olica del Norte,
 Av. Angamos 0610, Antofagasta, Chile\\
 $^2$Instituto de F\'{\i}sica La Plata
(CONICET--UNLP),
Instituto de F\'{\i}sica La Plata-CONICET\\ Fac.\ de Ciencias Exactas,
Universidad Nacional de La Plata, C.C.~67, 1900 La Plata, Argentina\\
 $^3$ Facultad de Ciencias Exactas, National
University La Pampa, Peru y Uruguay, Santa Rosa, La Pampa,
Argentina\\ }

\date{\today}

\begin{abstract}
\nd We discuss phase-space delocalization for the rigid rotator
within a semiclassical context by recourse to the Husimi
distributions of both the linear and the $3D-$anisotropic
instances. Our treatment is based upon the concomitant  Fisher
information measures. The pertinent Wehrl entropy is also
investigated in the linear case.
\end{abstract}
 \pacs{03.67.-a, 05.30.-d, 31.15.Gy}
 \keywords{Quantum Information, Quantum Statistical Mechanics, Semiclassical
methods.}

 \maketitle
 \newpage

\section{Introduction}

\nd   We will be concerned here with
 the semiclassical description of  the rotational dynamics of
  molecular systems. A pioneer effort in this sense is that of
  Morales {\it et al.} in Ref.~\cite{Morales}, who studied
   the connection between such dynamics
and coherent states~\cite{Glauber,Klauder}.

\nd   The coherent states formulation is not unique and several
authors have developed alternative descriptions for molecular
quantum systems. We have chosen two traditional formulations for
coherent states in order to discuss the semiclassical dynamics of
molecular rotational systems. One of them is devoted to two
dimensional case \cite{Sakurai} and the other  to the three
dimensional one \cite{Morales}.

\nd   In Refs. \cite{Casida,Harriman} the authors discuss, among
other things, the advantages of the Husimi distribution for the
interpretation of the electronic structure in   hydrogen and
nitrogen atoms. They suggested  that
 their work may be extended to the molecular instance.
 Following this suggestion we address here the  simplest applicable model,
i.e., the rigid rotator. Its usefulness for describing diatomic
molecules is well-known~\cite{pathria1993}.

\nd    Delocalization is an ``energetically favorable" process,
since it distributes the wave of function over a volume  greater
than the size of the sample.  Thus, the net energy of the molecule
is lowered, which results in resonance-stabilization. The
celebrated
 Fisher information $\mathcal{I}$ \cite{Frieden,roybook,roybook2,Pennini1}
 can be related to the
delocalization measure as follows: if we take a wave package with
a standard deviation $\sigma$ the Fisher information is given by
$\mathcal{I} \geq 1/\sigma^2$, thus, in this sense, we realize
that the Fisher information is a quite sensitive indicator of the
delocalization of the wave
package~\cite{Frieden,roybook,roybook2}.

\nd    The rigid rotator is a system of a single particle whose
quantum spectrum of energy is exactly known. Therefore, the study of
typical thermodynamic properties can be analytically
derived\cite{ullah}. Applications lead to the treatment of important
aspects of molecular systems\cite{arranz}.

\nd    The paper is organized as follows: In Section \ref{Linear}
we explore the linear rigid rotator. We write the probability of
finding a quantum state in a coherent state that is used to obtain
an explicit expression for the (i) Husimi distribution, (ii) Wehrl
entropy,  and (iii) Fisher Information. These results are of help
in Section~\ref{3D}, where  we discuss a model for the three
dimensional rigid rotator. Finally, in Section~\ref{remarks}, we
draw some conclusions.

\section{Linear rigid rotator}\label{Linear}
\nd We start the present  study by exploring a simple model, the
linear rigid rotator (LRR), based on the excellent discussion concerning
the coherent states for  angular momenta given in
Ref.~\cite{Nieto}. The LRR-hamiltonian
writes~\cite{pathria1993}

\be \hat H=\frac{\hat{L}^2}{2 I_{xy}}, \ee where
$\hat{L}^2=\hat{L}_x^2+\hat{L}_y^2$ is the angular momentum
operator and $I_x$ and $I_y$ are the associated moments of inertia.
We have assumed that $I_{xy}\equiv I_x=I_y$. Calling
${|I,K\rangle}$ the set of $H$-eigenstates, we recall that
 they verify the relations

\ben
\hat{L}^2 |IK\rangle\ &=&I(I+1)|IK\rangle\,\cr
\hat{L}_z |IK\rangle\ &=&K |IK\rangle,
\een with $I=0,1,2\ldots,$ for $ -I\leq K\leq I$, the eigenstates'
energy spectrum being given by

\be \varepsilon_I=\frac{I(I+1)\hbar^2}{2 I_{xy}}, \ee where for
simplicity we take  $\hbar=1$.

 \nd   We
construct the lineal coherent states for the rigid rotator using
 Schwinger's oscillator model of angular
momentum~\cite{Sakurai,Schwinger} as

\be |I K\rangle=
\frac{(\hat{a}_+^\dag)^{I+K}(\hat{a}_-^\dag)^{I-K}}{\sqrt{(I+K)!(I-K)!}}|0\rangle,
\label{lm} \ee with $\hat{a}_+$ and $\hat{a}_-$ the pertinent
creation and annihilation operators, respectively, and
$|0\rangle\equiv|0,0\rangle$  the vacuum state. The states $|I
K\rangle$ are  orthogonal and satisfy the closure relation, i.e.,
\be \langle I^{'} K^{'}|I K\rangle=\delta_{I^{'},I} \label{ortonorm}
\delta_{K^{'},K} ,\ee \be \sum_{I=0}^{\infty}\sum_{K=-I}^{I}|I
K\rangle\langle I K|=\hat 1. \label{clausura} \ee

\nd Since we deal with two degrees of freedom the ensuing
 coherent states are of the
tensorial product form (involving $|z_1\rangle$ and $|z_2
\rangle$) \cite{Nieto}

\be |z_1 z_2 \rangle=|z_1 \rangle\otimes| z_2\rangle, \ee
 where
 \be
\hat{a}_+|z_1 z_2 \rangle=z_1|z_1 z_2\rangle\label{a1}, \ee \be
\hat{a}_-|z_1 z_2\rangle=z_2|z_1 z_2\rangle.\label{a2} \ee
Therefore, the coherent state $|z_1 z_2 \rangle$ writes \cite{Nieto}

\be |z_1 z_2\rangle=e^{-\frac{|z|^2}{2}}e^{z_1
\hat{a}_+^\dag}e^{z_2 \hat{a}_-^\dag}|0\rangle, \label{z0} \ee
with

 \ben
|z_1\rangle&=&e^{-\frac{|z_1|^2}{2}}\,e^{z_1\hat{a}_+^\dag}|0\rangle ,\\
|z_2\rangle&=&e^{-\frac{|z_2|^2}{2}}\,e^{z_2\hat{a}_-^\dag}|0\rangle.
\een We have introduced the convenient notation

 \be |z|^2=|z_1|^2+|z_2|^2.\label{z12} \ee

\nd Using Eqs.~(\ref{lm}) and~(\ref{z0}) we easily
 calculate $\langle I K|z_1 z_2\rangle$ and,  after a bit of algebra,
find

\be \langle I K|z_1
z_2\rangle=e^{-\frac{|z|^2}{2}}\,\frac{z_1^{n_+}}{\sqrt{n_+!}}\,\frac{\,z_2^{n_-}}{\sqrt{n_-!}}
\ee where $n_+=I+K$ and $n_-=I-K$. Therefore, the probability of
observing the state $|I K\rangle$ in the coherent state $|z_1
z_2\rangle$ is of the form

\be |\langle
I K|z_1 z_2\rangle|^2=e^{-|z|^2}\,
\frac{z_1^{2n_+}}{n_+!}\,\frac{\,z_2^{2n_-}}{n_-!}\label{mod2}.
\ee

\subsection{Husimi distribution}\label{Sub_Husimi}
\nd Following the procedure developed by Anderson {\it et
al.}~\cite{PRD2753_93},  we  can readily calculate the Husimi
distribution~\cite{Husimi}, which is defined as

\be  \label{X1} \mu(z_1,z_2)=\langle z_1,z_2 \vert \hat\rho \vert
z_1,z_2\rangle, \ee
where the density operator is

 \be \label{X2} \hat
\rho=Z_{2D}^{-1}\,\exp{(-\beta \mathcal{\hat H})},   \ee and
$\beta=1/k_BT,$ $k_B$ is the Boltzmann's constant and $T$ the
temperature. The form of the rotational partition function $Z_{2D}$
is given in Ref.~\cite{pathria1993}

\be Z_{2D}=\sum_{I=0}^{\infty} (2I+1)\, e^{-I(I+1)\frac{\Theta}{T}},
\label{Z2D}\ee with $\Theta=\hbar^2/(2I_{xy} k_B)$. In the present
context, speaking of the ``trace operation" entails performing the
sum $ \tr \equiv \sum_{I=0}^{\infty}\sum_{K=-I}^{I}.$

\begin{figure}[h]
   \centering
  \resizebox{0.9\columnwidth}{!}{%
\includegraphics{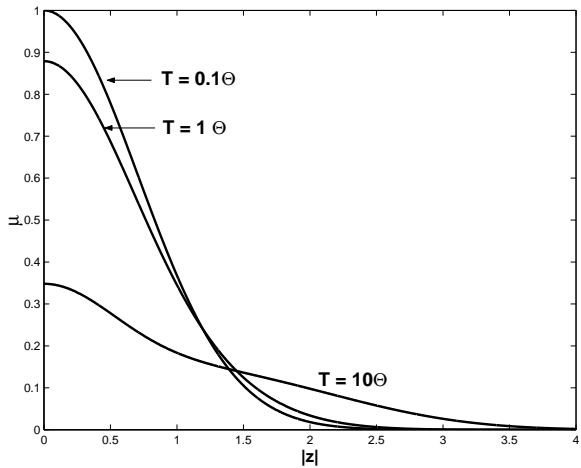}}
        \caption{It is depicted the Husimi function $\mu(z_1,z_2)$ as a function of $|z|$ for several values of the
        temperature, ($T=0.1\Theta, \Theta, 10\Theta$) for the linear rotator.
        The behavior of the Husimi function looks like Gaussian distribution.
        The peak of the distribution increases as the temperature decreases.}
   \label{fig1}
\end{figure}

\nd Inserting the closure  relation into Eq.~(\ref{X1}), and using
Eq.~(\ref{mod2}),
 we finally get our Husimi distributions in the fahion

\be \mu(z_1,z_2)=e^{-|z|^2}\,
\frac{\sum_{I=0}^{\infty}\,\frac{|z|^{4I}}{(2I)!}\,e^{-I(I+1)\,\frac{\Theta}{T}}}{\sum_{I=0}^{\infty}\,
(2I+1) \, e^{-I(I+1)\frac{\Theta}{T}}}.\label{mur} \ee \nd It is
easy to show that this distribution is  normalized to unity

\be \int\,\frac{\mathrm{d}^2z_1}{\pi}
\frac{\mathrm{d}^2z_2}{\pi}\,\mu(z_1,z_2)=1,\label{normaliz}
 \ee
where $z_1$ and $z_2$ are given by Eqs. (\ref{a1}), (\ref{a2}),
and (\ref{z12}). Note that we  must deal with  the binomial
expression $(|z_1|^2+|z_2|^2)^{4 I}$ firstly and then integrate
over the whole complex plane (in two dimensions) in order to
verify the normalization
 condition~(\ref{normaliz}). The differential element of area in the $z_1$($z_2$) plane is
$\mathrm{d}^2z_1=\mathrm{d}x\mathrm{d}p_x/2\hbar$
 ($\mathrm{d}^2z_2=\mathrm{d}y\mathrm{d}p_y/2\hbar$)~\cite{Glauber}.
Moreover, we have  the phase-space relationships

\begin{subequations}
\be |z_1|^2=\frac14
\left(\frac{x^2}{\sigma_x^2}+\frac{p_x^2}{\sigma_{
p_x}^2}\right),\label{zp1} \ee \be |z_2|^2=\frac14
\left(\frac{y^2}{\sigma_y^2}+\frac{p_y^2}{\sigma_{p_y}^2}\right),\label{zp2}
\ee
\end{subequations}
where $\sigma_x\equiv\sigma_y=\sqrt{\hbar/2m\omega}$ and
$\sigma_{p_x}\equiv\sigma_{p_y} =\sqrt{\hbar m\omega/2}$. In
Fig.~\ref{fig1}, we depict  the behavior of the Husimi distribution
$\mu(z_1,z_2)$ as a function of $|z|$ at fixed temperature. The
profile of the Husimi function is similar to Gaussian distribution.
\subsection{Wehrl entropy and Fisher Information}
\label{Sub_WehrlFish}

\nd The Wehrl entropy is a semiclassical measure of
localization~\cite{Wehrl}. So is Fisher's one \cite{Pennini1}.
 The Wehrl measure is simply a logarithmic Shannon
measure built up with Husimi distributions. For the present
bi-dimensional model this entropy is of the form

\be
\mathcal{W}=\int\,\frac{\mathrm{d}^2z_1}{\pi}\frac{\mathrm{d}^2z_2}{\pi}\,\mu(z_1,z_2)\,\ln
\mu(z_1,z_2), \ee where $\mu(z_1,z_2)$ is given by
Eq.~(\ref{mur}). The so-called phase-space, shift-invariant Fisher
measure \cite{Pennini1} is a particular instance of the general
Fisher-one \cite{roybook,roybook2},
 that can be regarded as an (also semiclassical)
  counterpart of Wehrl entropy \cite{Pennini1}.
 Extending now to the present $2D-$case the ideas
 developed in Ref.~\cite{Pennini1} for the case of the
$1D-$harmonic oscillator, we  define the (phase-space) shift
invariant Fisher measure in the fashion

 \be \mathcal{I}=\int
\,\frac{\mathrm{d}^2z_1}{\pi}\frac{\mathrm{d}^2z_2}{\pi}\,\mu(z_1,z_2)\,\mathcal{A},
\ee with

\begin{equation} \mathcal{A}=\sum_{\Lambda=\{x,p_x,y,p_y\}}\,\sigma_{\Lambda}^2\left[\frac{\partial \ln
  \mu(z_1,z_2)}{\partial \Lambda}\right]^2, \label{AFisher}
\end{equation}
where we have introduced a simplified in which the index $\Lambda$
successively takes the values $x,p_x,y$, and~$p_y$, respectively.
It is easy to prove that the quantity $\mathcal{A}$ has the
following form --see details in appendix~\ref{Apendix}--

\be \mathcal{A}=\eta(z_1,z_2)^2, \ee where

\be
\eta(z_1,z_2)=\frac{\sum_{I=0}^\infty\,\left[\frac{|z|^{4I-1}}{(2I-1)!}-
\frac{|z|^{4I+1}}{(2I)!}\right]
\,e^{-I(I+1)\Theta/T}}{\sum_{I=0}^\infty\,\frac{|z|^{4I}}{(2I)!}\,e^{-I(I+1)\Theta/T}}.
\label{eta} \ee Therefore, the corresponding Fisher measure
acquires the simpler appearance

\be \mathcal{I}= \int \frac{\mathrm{d}^2z_1}{\pi}
\frac{\mathrm{d}^2z_2}{\pi}\,\mu(z_1,z_2)\,\,\eta(z_1,z_2)^2,
\ee
i.e.,
\be
  \mathcal{I} \equiv \langle \eta(z_1,z_2)^2\rangle,
\ee where with the notation

\be \langle \mathcal{G}\rangle=\int
\frac{\mathrm{d}^2z_1}{\pi}
\frac{\mathrm{d}^2z_2}{\pi}\,\mu(z)\,\mathcal{G},
\ee we allude to the {\it semi-classical expectation value} of~$\mathcal{G}$.

\begin{figure}[h]
   \centering
   \resizebox{0.9\columnwidth}{!}{%
\includegraphics{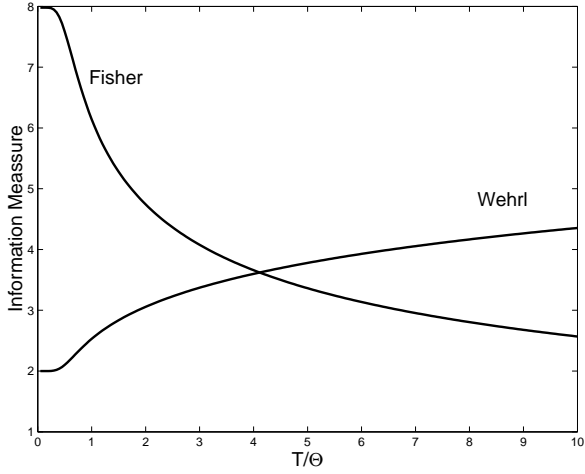}}
        \caption{The trend of the Fisher information measure (delocalization) ($\mathcal{I}$) in comparison
        with the Wehrl entropy ($\mathcal{W}$) as a function of the temperature are shown for the linear rotator.
        We see that, if the temperature increases, then the delocalization decreases while Wehrl entropy increases.}
  \label{fig2}
\end{figure}

\nd In Fig.~\ref{fig2} we plot both Fisher's information and
Wehrl's entropy as a function of $T/\Theta$. They behave in
different manner. If the temperature $T\rightarrow 0,$  Fisher's
information measure (inverse-delocalization) takes its maximum
value and Wehrl's its minimum.  This behavior is reversed for high
temperatures, with the degree of delocalization becoming larger
and larger.

\section{Rigid rotator in three dimensions}\label{3D}
\nd In  the present
 section  we consider a
more general problem, the $3D-$rigid rotator model, whose
hamiltonian writes~\cite{Morales}

\be \hat H=\frac{\hat{L}_x^2}{2 I_x} + \frac{\hat{L}_y^2}{2 I_y} +
\frac{\hat{L}_z^2}{2 I_z},\ee where $I_x$, $I_y$, and $I_z$ are
the associated moments of inertia. A complete set of rotor
eigenstates is $\{|IMK\rangle\}$. The following relations apply

\ben
\hat{L}^2 |IMK\rangle\ &=&I(I+1)|IMK\rangle\,\nonumber\\
\hat{L}_z |IMK\rangle\ &=&K |IMK\rangle\\\
\hat{J}_z |IMK\rangle\ &=&M |IMK\rangle,\,\nonumber \een where $I=0,
\ldots, \infty, -I\leq K \leq I,$ and $-I\leq M \leq I$. The states
$|IMK\rangle$ satisfy orthogonality and  closure
relations~\cite{Morales}

 \be \langle I^{'}M^{'}K^{'}|IMK\rangle=\delta_{I^{'},I}
\delta_{M^{'},M}  \delta_{K^{'},K} \label{ortonorm2}\ee \be
\sum_{I=0}^{\infty}\sum_{M=-I}^{I}\sum_{K=-I}^{I}|IMK\rangle\langle
IMK|=\hat 1. \label{clausura2} \ee
 If we take
$\hat{L}^2=\hat{L}_x^2+\hat{L}_y^2+\hat{L}_z^2$ and assume axial symmetry, i.e.,
$I_{xy}\equiv I_x=I_y$, we can recast the
hamiltonian as

 \be
\hat H=\frac{1}{2I_{xy}}\left[\hat{L}^2 + \left( \frac{I_{xy}}
{I_z}-1\right)\hat{L}_z^2 \right],\ee where $\hat{L}^2$ is the
angular momentum operator and $\hat{L}_z$ is its projection on the
rotation axis $z$. The concomitant spectrum of energy becomes

\be \varepsilon_{I,K}=\frac{\hbar^2}{2 I_{xy}}\left[ I(I+1)+\left(
\frac{I_{xy}} {I_z}-1\right)\,K^2 \right], \ee where
$I=0,1,2,\cdot\cdot\cdot$ and it represents the eigenvalue of the
angular momentum operator $\hat{L}^2$, the numbers
$m=-I,\cdot\cdot\cdot,-1,0,1,\cdot\cdot\cdot,I$ represent the
projections on the intrinsic rotation axis of the rotor. All states
present a degeneracy given by $(2I+1)$. The parameters
$I_x=I_y\equiv I_{xy}$ and $I_z$ are the inertia momenta. Several
geometrical cases are characterized through the $I_{xy}/I_z$ ratio.
For instance, the marginal value $I_{xy}/I_z=1$ corresponds to the
spherical rotor. Limiting cases can be considered; this is,
$I_{xy}/I_z=1/2$ and  $I_{xy}/I_z\rightarrow \infty$ that correspond
to the extremely oblate and prolate cases, respectively.

\subsection{Coherent states}
\nd  In order to obtain the Husimi distribution for this problem
we need first of all to construct the associated coherent states,
as already discussed   by Morales {\it et al.}
 in Ref.~\cite{Morales}. Introduce first the auxiliary quantity

\be X_{I,M,K}=\sqrt{I!(I+M)!(I-M)!(I+K)!(I-K)!}, \ee
and then write~\cite{Morales}

{\small
\be
 |z_1 z_2 z_3\rangle = e^{-\frac{|u|^2}{2}} \sum_{IMK}
\frac{[(2I)!]^2z_1^{(I+M)}z_2^I z_3^{(I+K)}}{ X_{I,M,K}}
|IMK\rangle, \label{IKM} \ee}where the following supplementary
variable were introduced~\cite{Morales}

 \be |u|^2=|z_2|^2 (1+|z_1|^2)^2 (1+|z_3|^2)^2. \ee

\subsection{Husimi function, Wehrl entropy, and Fisher measure}
 \nd Using now Eq.~(\ref{IKM}) we find
{\small
\be
|\langle IMK | z_1z_2z_3 \rangle|^2
=\frac{e^{-|u|^2}}{X_{I,M,K}^2}[(2I)!]^2|z_1|^{2(I+M)}|z_2|^{2I}
|z_3|^{2(I+K)}
\ee}and determine that, in this case, the rotational partition
function reads

\be
 Z_{3D}=\sum_{I=0}^{\infty}\,\sum_{K=-I}^{I}\,
 \sum_{M=-I}^{I}\,e^{-\beta \varepsilon_{I,K}},
\ee
i.e.,
\be
 Z_{3D}=\sum_{I=0}^{\infty}\,(2I+1)\,e^{-I(I+1)\frac{\Theta}{T}}
\sum_{K=-I}^{I}\,e^{-\left(\frac{I_{xy}}
{I_z}-1\right)K^2\frac{\Theta}{T}}. \label{Z3D}\ee It is convenient
to remark that if we take one of limiting cases, this the extremely
prolate, $I_{xy}/I_z\rightarrow\infty$, the only term that survives
in the right sum of the right side in Eq.(\ref{Z3D}) is that for
$K=0$ and all terms for $K\neq 0$ vanish; in such particular case
$Z_{2D}$ is recovered from $Z_{3D}$.

The pertinent Husimi distribution then becomes \be \mu(z_1,z_2,z_3)
=\frac{e^{-|u|^2
}}{Z_{3D}}\sum_{I=0}^{\infty}\,\frac{(2I)!}{I!}\,|v|^{2I}\,e^{-I(I+1)\frac{\Theta}{T}}\times
g(I)\label{Husi3D} \ee where \be
g(I)=\sum_{K=-I}^{I}\frac{|z_3|^{2(I+K)}}{(I+K)!(I-K)!}\,
e^{-\left(\frac{I_{xy}} {I_z}-1\right)K^2\frac{\Theta}{T}}, \ee with
\begin{subequations}
\be
|v|^2=(1+|z_1|^2)^{2} |z_2|^{2},
\ee
\be
|u|^2=|v|^2(1+|z_3|^2)^2.
\ee
\end{subequations}
We can easily verify that $\mu(z_1,z_2,z_3)$ is  normalized in the
fashion

\be\int\,\mathrm{d}\Gamma\,\mu(z_1,z_2,z_3)=1,\label{normaliz3}\ee
where $\mathrm{d}\Gamma$ is the measure of integration given by~\cite{Morales}

\begin{eqnarray}
 \mathrm{d}\Gamma&=&\mathrm{d}\tau
 \bigg\{4[(1+|z_1|^2)(1+|z_3|^2)]^4|z_2|^4-\nonumber\\&&-8[(1+|z_1|^2)(1+|z_3|^2)]^2|z_2|^2+1\bigg\}
 \label{measure3D}
\end{eqnarray}
with
\be
\mathrm{d}\tau=\frac{\mathrm{d}^2z_1}{\pi}\frac{\mathrm{d}^2z_2}{\pi}\frac{\mathrm{d}^2z_3}{\pi},
\ee

\nd where, of course, in this case we have three degrees of freedom.

We compute then, the Wherl entropy in the form

\be
\mathcal{W}=\int \,\mathrm{d}\Gamma\,\mu(z_1,z_2,z_3)\,\ln \mu(z_1,z_2,z_3),
\ee
and explicitly the Fisher measure as follows

\be
\mathcal{I}=\int\,\mathrm{d}\Gamma\,\mu(z_1,z_2,z_3)\,\mathcal{A}_{3D}\label{Fisher3D}
\ee where in this case we define the quantity $\mathcal{A}_{3D}$ in
three dimensions as

\be
\mathcal{A}_{3D}=\sum_{\Lambda=\{x,p_x,y,p_y,z,p_z\}}\,\sigma_{\Lambda}^2\left[\frac{\partial \ln
  \mu(z_1,z_2,z_3)}{\partial {\Lambda}}\right]^2,\label{A3DFisher}
\ee
with the phase space relationships (\ref{zp1}), (\ref{zp2}) and

\be |z_{3}|^2=\frac14
\left(\frac{z^2}{\sigma_z^2}+\frac{p_z^2}{\sigma_{ p_z}^2}\right),
\ee where  $\sigma_z=\sqrt{\hbar/2m\omega}$ and
$\sigma_{p_z}=\sqrt{\hbar m\omega/2}$. In this instance
$\mathrm{d}^2z_3=\mathrm{d}z\mathrm{d}p_z/2\hbar$. So, after a bit
of algebra we arrive to

\begin{eqnarray}
\mathcal{I}&=&\int\,\mathrm{d}\Gamma\,\mu(z_1,z_2,z_3)\,\bigg\{\gamma^2(|z_1|^2|z_2|^2+\frac14(1+|z_1|^2)^2)
+\nonumber\\&&+4|u|^2|z_3|^2\bigg\},
\end{eqnarray}
i.e., \be \mathcal{I}=\langle
\gamma^2(|z_1|^2|z_2|^2+\frac14(1+|z_1|^2)^2)+4|u|^2
|z_3|^2\rangle \ee where

\begin{eqnarray}
\gamma&=&\frac{-(1+|z_3|^2)^2\sum_{I=0}^{\infty}\,\frac{2(2I)!}{I!}\,|v|^{2I+1}\,e^{-I(I+1)\frac{\Theta}{T}}\times g(I)}{\sum_{I=0}^{\infty}\,\frac{(2I)!}{I!}\,|v|^{2I}\,e^{-I(I+1)\frac{\Theta}{T}}\times g(I)}
+\nonumber\\&&+\frac{\sum_{I=0}^{\infty}\,\frac{2(2I)!}{(I-1)!}\,|v|^{2I-1}\,e^{-I(I+1)\frac{\Theta}{T}}\times g(I)}{\sum_{I=0}^{\infty}\,\frac{(2I)!}{I!}\,|v|^{2I}\,e^{-I(I+1)\frac{\Theta}{T}}\times g(I)}.
\end{eqnarray}

In the special instance  $I_{xy}/I_z=1$, that corresponds to the
spherical rotor, we can explicitly obtain

\be \mu(z_1,z_2,z_3)=e^{-|u|^2 }\,\frac{\sum_{I=0}^{\infty}\,\frac{
|u|^{2I}}{I!}\,e^{-I(I+1)\frac{\Theta}{T}}}{\sum_{I=0}^{\infty}\,(2I+1)^2\,e^{-I(I+1)\frac{\Theta}{T}}
}. \ee

\nd Having the Husimi functions the Wehrl entropy is
straightforwardly computed.

 \begin{figure}[h]
   \centering
   \resizebox{\columnwidth}{!}{%
\includegraphics{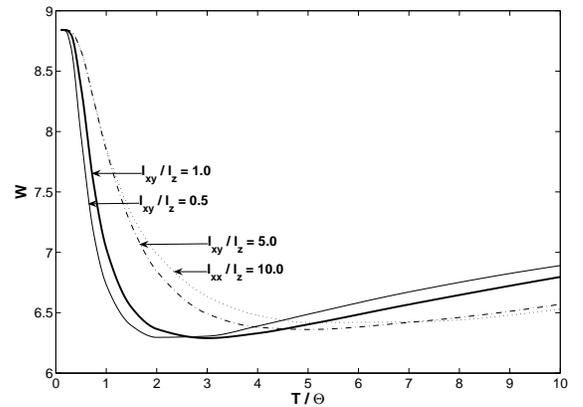}}
        \caption{A family of Wehrl entropy $\mathcal{W}$ as a function of $T/\Theta$ is depicted for the 3D
        anisotropic rigid rotator for several values of the anisotropy ($I_{xy}/I_z = 1/2,1,5,10)$. The trend of the
        Wehrl entropy is not very similar to the linear case, it has a maximum in $T/\Theta=0$,
        a minimum in $T/\Theta \geq 2$, whose exact value depends on the ratio $I_{xy}/I_z$.}
  \label{wehrl3D}
\end{figure}

 In figure \ref{wehrl3D} we compare
the Wehrl entropy $\mathcal{W}$, as a function of $T/\Theta$, for
several values of $I_{xy}/I_z$; this is, the extremely oblate
rotator $I_{xy}/I_z= 1/2$ (e.g., $CHCl_3$ and $C_6H_6$), the prolate
rotator $I_{xy}/I_z= 5, 10$ (e.g., $CH_3Cl$ and $PCl_5$) and the
spherical case $I_{xy}/I_z= 1$. Unfortunately, from the 3D
formulation of coherent states is not possible to recover the form
of the Wehrl entropy, in the same way as occurs with the partition
function from the Eq.(\ref{Z3D}), which is conveniently lead to the
form of the Eq.(\ref{Z2D}) in the limiting case of the extremely
prolate rotator $I_{xy}/I_z\rightarrow \infty$. It could be due to a
note given in Ref.\cite{Morales}, the present version of coherent
states formulation is weak  because the measure from the
Eq.(\ref{measure3D}) is not positive.

\section{Concluding remarks}\label{remarks}
\nd We have concentrated our effort on
 the study of the semiclassical behavior of the rigid rotator and
 have obtained in  analytical fashion
  the form of the Husimi distribution for two cases, namely,
  the  linear and the axially symmetric  rigid rotator. As it is
  expected, the linear case is obtained as a particular case from the
  formulation in three dimensions.
Also, we have obtained an analytical expression of the
shift-invariant Fisher measure built up with Husimi distribution,
for the rigid rotator model, concluding that Fisher measure is
better indicator of the delocalization than Wherl entropy for this
model. The present study could motivate other specialists to improve
the present formulation of the coherent states in order to recover
all quantities in two dimensions from a formulation in three
dimensions.

\vspace{0.5cm} \nd {\bf Acknowledgment} S. Curilef and F. Pennini
would like to thank partial financial support by FONDECYT, grant
1051075.
\newpage


\section{Appendix: A bit of algebra}
\label{Apendix}

First of all, we carry out the differentiation the Husimi
distribution~(\ref{mur}) with respect to the variable $x$,
obtaining the following result

\be
 \frac{\partial \ln \mu(z_1,z_2)}{\partial x}=2 \eta(z_1,z_2) \frac{\partial |z|}{\partial x}
\ee
where the quantity $\eta(z_1,z_2)$ was defined in Eq.~(\ref{eta}). Moreover, from Eqs.~(\ref{zp1}) and (\ref{zp2}) we have

\be
\frac{\partial |z|}{\partial x}=\frac{x}{4 |z| \sigma_x^2},
\ee
and we are lead to

\be
\frac{\partial \ln \mu(z_1,z_2)}{\partial x}= \frac{\eta(z_1,z_2) x}{2 |z| \sigma_x^2}.
\ee

We arrive to a similar expression differentiating with respect to
$p_x$,

\be
\frac{\partial \ln \mu(z_1,z_2)}{\partial p_x}= \frac{\eta(z_1,z_2) p_x}{2 |z| \sigma_{p_x}^2}.
\ee

Analogous expressions are obtained replacing $x$ and $p_x$ for $y$
and $p_y$. Finally, substituting these results into~Eq.
(\ref{AFisher}) we thus arrive to

\be
\mathcal{A}=\eta(z_1,z_2)^2.
\ee


\end{document}